\def\year{2022}\relax
\documentclass[letterpaper]{article} 
\usepackage{aaai22}  
\usepackage{times}  
\usepackage{helvet}  
\usepackage{courier}  
\usepackage[hyphens]{url}  
\usepackage{graphicx} 
\usepackage{multirow}
\usepackage{natbib}  
\usepackage{caption} 
\DeclareCaptionStyle{ruled}{labelfont=normalfont,labelsep=colon,strut=off} 
\usepackage{algorithm}
\usepackage{algorithmic}

\usepackage{newfloat}
\usepackage{listings}
\floatstyle{ruled}
\newfloat{listing}{tb}{lst}{}
\floatname{listing}{Listing}
\title{Single-Read Reconstruction for DNA Data Storage Using Transformers}
\author {
    Yotam Nahum, \equalcontrib\textsuperscript{\rm 1}
    Eyar Ben-Tolila,\equalcontrib \textsuperscript{\rm 2}
    Leon Anavy \textsuperscript{\rm 3}
}
\affiliations {
    \textsuperscript{\rm 1} Data Science Institute, Reichman University (IDC Herzliya)\\
    \textsuperscript{\rm 2}  School  of  Electrical  and  Computer  Engineering,  Ben-Gurion  University  of  the  Negev\\
    \textsuperscript{\rm 3}  Efi Arazi School of Computer Science, Reichman University (IDC Herzliya)\\    
    yotamnahum@gmail.com, eyarbento@gmail.com, leonanavy@gmail.com
}

\usepackage{bibentry}

\begin{document}

\maketitle

\begin{abstract}
As the global need for large-scale data storage is rising exponentially, existing storage technologies are approaching their theoretical and functional limits in terms of density and energy consumption, making DNA based storage a potential solution for the future of data storage. Several studies introduced DNA based storage systems with high information density (petabytes/gram). However, DNA synthesis and sequencing technologies yield erroneous outputs. Algorithmic approaches for correcting these errors depend on reading multiple copies of each sequence and result in excessive reading costs. The unprecedented success of Transformers as a deep learning architecture for language modeling has led to its repurposing for solving a variety of tasks across various domains. In this work, we propose a novel approach for single-read reconstruction using an encoder-decoder Transformer architecture for DNA based data storage. We address the error correction process as a self-supervised sequence-to-sequence task and use synthetic noise injection to train the model using only the decoded reads. Our approach exploits the inherent redundancy of each decoded file to learn its underlying structure. To demonstrate our proposed approach, we encode text, image and code-script files to DNA, produce errors with high-fidelity error simulator, and reconstruct the original files from the noisy reads. Our model achieves lower error rates when reconstructing the original data from a single read of each DNA strand compared to state-of-the-art algorithms using 2-3 copies. This is the first demonstration of using deep learning models for single-read reconstruction in DNA based storage which allows for the reduction of the overall cost of the process. We show that this approach is applicable for various domains and can be generalized to new domains as well.
\end{abstract}

\begin{figure*}[htp]
    \centering
  \includegraphics[scale=0.8]{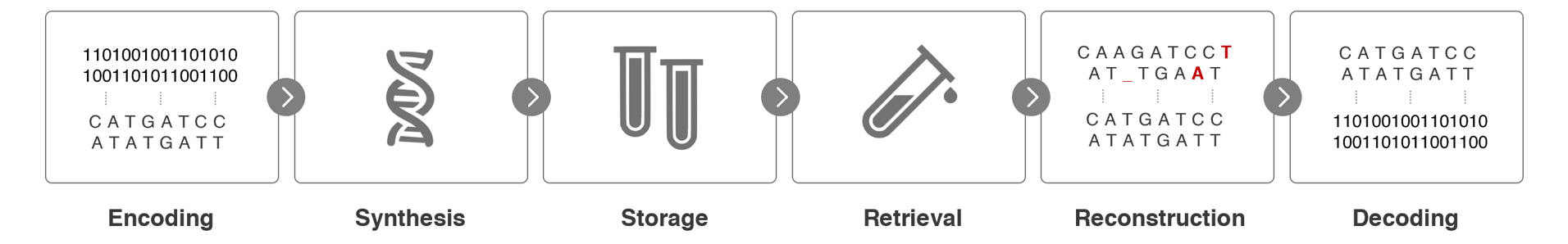}
  \caption{\textbf{DNA based storage and sequence reconstruction.} An information baring file is encoded into a set of DNA sequences. DNA molecules are synthesizes, stored and sequenced, generating an output set of noisy reads. The original sequences are reconstructed and decoded to retrieve the original stored message.} 
  \label{fig:dna_Storage_Pipeline}
\end{figure*}

\section{Introduction}
In recent years, the amount of digital data created and stored by humanity is rapidly growing. Current data storage technologies are limited in their ability to support the growing demand, and thus the need for alternative storage media is rising. The durability and high information density of DNA make it an appealing medium for future data storage systems \cite{Church2012,Goldman2013}. A single gram of DNA can potentially store 100s of petabytes and expected to last and remain readable for centuries. In comparison, standard storage media such as optical discs, hard drives and magnetic tapes,  only enable data lifetime of a few years with an information density which is lower by orders of magnitude \cite{chandak2020overcoming,char:19}. 

DNA based storage systems are comprised of three main components. First, the input information is encoded into a set of DNA sequences (\textit{i.e.} sequences over the four letter alphabet $\{A,C,G,T\}$). Next, the sequences are ``transmitted" over a DNA based channel including synthesis, storage, and sequencing. Finally, the sequencing output is reconstructed and decoded. Figure \ref{fig:dna_Storage_Pipeline} presents an overall view of DNA based storage.  
The encoding step involves adjustments to technological and biological constrains, such as adding error-correction codes to the sequences. Standard synthesis technologies generate thousands to millions of copies for every 100-200 base sequence \cite{char:19}. Moreover, all the synthesized DNA molecules are stored together in a single container as an unordered set. These properties, among others, impose different limitations on the encoding step including sequence length limitation, inedxing, controlled GC content, and run-length constrains.
The output of the sequencing process is a set of noisy sequences, usually referred to as reads, representing the synthetic DNA molecules.
Retrieving the information from this read set involves sequence reconstruction, ordering, and decoding.

Despite the promising potential, many challenges prevent DNA based storage from becoming a viable solution, even for archival purposes. The synthesis and sequencing processes are costly and result in erroneous reads that include insertion, substitution and deletion errors \cite{char:19}. Furthermore, as the number of sequenced molecules is considerably lower than the number of available molecules, only a small sample is sequenced, thus some sequences  are read multiple times while others might not be read at all. 
Some recent work on DNA based storage systems presented end-to-end solutions including molecular implementations \cite{Organick2018random,df:17,organick2020probing}. Others tackled certain components: error correction schemes \cite{nguyen2021capacity, lenz2020coding} or read clustering and sequence reconstruction algorithms \cite{ccc:19,Sabary:2020,chandak2019improved}. All the algorithms suggested for the reconstruction step, which is the task of predicting the original sequences from the noisy read set, use multiple copies for reconstruction. Obtaining enough sequences with sufficient read coverage requires excess reading which entails higher costs. For more information regarding DNA data storage, common challenges, and recent work, we refer the reader to \cite{Ceze2019}. 

During the past few years, Transformers \cite{vaswani2017attention} have become the most dominant architecture for many sequence-to-sequence tasks as it showed to be superior to methods that have been used before (mainly CNNs and RNNs). Although initially developed for language modelling tasks, Transformers are now widely used in other areas including audio processing \cite{dong2018speech} and image processing \cite{parmar2018image}. More recently Transformers also achieved significant impact on one of the cornerstones of computational biology, the problem of protein folding prediction \cite{jumper2021highly}.
A Transformer is a neural network with permutation equivalence composed of attention mechanisms and residual blocks. In the Transformer’s architecture, position embeddings are generally added into the embedding layer with token embedding, since Transformers cannot distinguish the input’s position. 

In this work, we focus on reconstructing the original DNA sequences from only a single noisy read of each sequence. We address the problem as a sequence-to-sequence task and use an ensemble of three encoder-decoder Transformer models for the reconstruction. Our model relies solely on self-supervised learning using noise injection on reads from the decoded file that are presumed to be error free with high probability \cite{lewis2019bart}. We evaluate our method on files from three different data domains, to which we simulate reads with realistic error rates. With average error rate reductions of $63\%$ (edit error), and $97\%$ (hamming error) using only a single read, our model outperforms state-of-the-art algorithms using 2-3 copies. Furthermore, the advantage of our method becomes more significant when tested over files with higher noise levels.

\textbf{Our contributions}: This work is the first to use a single-read for sequence reconstruction in DNA based storage. We demonstrate that our approach outperforms state-of-the-art algorithms that use two and three reads for reconstruction. Additionally, this work is the first attempt to incorporate deep learning models for sequence reconstruction in DNA based storage systems. Furthermore, our approach requires no training on external data, making it adjustable for the storage of different file types. Lastly, our approach represent a paradigm shift in sequence reconstruction as we focus on correcting the errors by learning the representation of the sequences rather than learning the error patterns.   

\section{Problem Definition}
The DNA channel takes a binary input message $\mathbf{m}$ and encodes it into a sequence of DNA codewords which is broken into $n$ sub-sequences of length $l$ each, $\mathbf{x_1},\mathbf{x_2},...\mathbf{x_n}$ where $\mathbf{x_i}\in \Sigma^l$ and $\Sigma = \{A,C,G,T\}$. For each $l$-bases sequence $\mathbf{x}$, $t>0$ independent copies are transmitted through a Deletion-Insertion-Substitution (DIS) channel yielding $\mathbf{y_1},\mathbf{y_2},...\mathbf{y_t}$. These, in turn are being decoded back to receive the binary output message $\hat{\mathbf{m}}$.

A general \emph{DNA reconstruction} algorithm is a mapping $R:\left (\Sigma^* \right)^t \rightarrow \Sigma^*$ which receives $\mathbf{y_1},\mathbf{y_2},...\mathbf{y_t}$ and produces $\hat{\mathbf{x}}$. The goal in a DNA reconstruction problem is to minimize either the edit distance or the Hamming distance between the original sequence $\mathbf{x}$ and the inferred sequence $\hat{\mathbf{x}}$, $d_e(\hat{\mathbf{x}},\mathbf{x})$ and $d_H(\hat{\mathbf{x}},\mathbf{x})$ respectively (See definitions below).

In this work we tackle the case for which $t=1$. For that purpose we generalize the problem definition by taking into account the entire message transmitted through the DNA channel. Formally, a reconstruction algorithm will be a mapping $R:\left (\Sigma^* \right)^n \rightarrow \left (\Sigma^l \right)^n$. That is, given a single noisy copy $\mathbf{y}$ for every sequence $\mathbf{x}$ we infer $\hat\mathbf{x}$. Moreover, the algorithm is aware of the overall encoding process of $\mathbf{m}$ to $\left ( \mathbf{x} \right )^n$ and can also enforce constraints on this encoding. This is sometimes referred to as \emph{coded reconstruction}. The purpose remains to minimize either $d_e(\hat{\mathbf{x}},\mathbf{x})$ or $d_H(\hat{\mathbf{x}},\mathbf{x})$.

\subsection{Performance Measures}
The edit distance between two sequences $x,y \in \Sigma^n$, denoted $d_e(x,y)$, is the minimum number of insertions, deletions and substitutions required to transform x into y.
The Hamming distance between $x$ and $y$, denoted $d_H(x,y)$ when $|x|=|y|$, is the number of substitutions required to transform $x$ into $y$.

\section{Related Work}
\subsection{DNA Based Storage Error Characterization}
\citeauthor{heckel2019characterization} studied the errors in a DNA storage channel by analyzing experimental data. They characterized deletion, insertion and substitution rates and the conditional error probabilities. 
\citeauthor{organick2020probing} examined the limits of DNA dilution and extensive PCR amplification on the overall reliability of DNA based storage systems. They showed that this is affected not only by the quality of the DNA synthesis and sequencing process, but also by the overall size of the files being stored together. 

\citeauthor{sabary2021solqc} performed an extensive analysis of the error patterns using experimental data from various DNA synthesis and sequencing technologies. They analyzed the error rates, including base specific and position specific rates. They also published SOLQC, a tool designed for the analysis of the error patterns in synthetic DNA libraries.

\subsection{Reconstruction Algorithms}

\citeauthor{BMA:18} extended the Bitwise Majority Alignment reconstruction algorithm suggested by \citeauthor{batu2004reconstructing} to support the DIS channel of DNA based storage. They aligned all reads by considering a majority vote per symbol, and for any read that its current symbol did not match the majority voting, they examined the next two symbols to determine the error type. \citeauthor{Organick2018random} used this reconstruction algorithm for a large scale implementation of an end-to end DNA based storage system.

\citeauthor{Sabary:2020} presented several reconstruction algorithms for DNA based storage channel, designed to minimize the edit distance between the original sequence and the estimated sequence. Their algorithms examine the entire read and use the shortest common super-sequence and the longest common sub-sequence of a given set of reads. The suggested algorithms achieved reduced edit-error rates compared to previously published algorithms while using less copies. However, their results are based on using 10 reads or more of each sequence, and in the case of a single read, the algorithm would simply return the read as is.

\subsection{Sequence Reconstruction Using Deep Learning}
\citeauthor{chandak2020overcoming} integrated classical error-correction algorithms and a recurrent neural network (RNN) to reduce error rates introduced in nanopore basecalling in DNA based storage. They significantly reduced  error rates achieving a $3\times$lower reading costs than compared works. However, their approach is specific to nanopore sequencing as it operates directly on the raw signal of the sequencer rather than on the obtained reads. 

\citeauthor{Pan2021.02.22.432304} showed how data can be stored in DNA both in the strand and in its backbone structure, to create a rewriteable DNA storage system and used ML for correction of stored image files.

In a work parallel to this work, \citeauthor{Barlev2021DL}. presented a scalable and robust end-to-end solution for DNA storage systems that is based on coding theory and DNN. To retrieve the erroneous data, their method exploits the error behavior of the DNA storage channel, instead of the inherent redundancy of the files.

\section{DNA Encoder-Decoder Transformer}
In this section we describe our model. Essentially, it is a sequence-to-sequence model that performs error correction by mapping a corrupted sequence to the original one.
\subsection{Pre-Processing}
\subsubsection{Overlapping k-mer Sequence Representation}
Each input sequence $\mathbf{s} \in \Sigma^n$ is represented using an overlapping k-mer representation. The k-mer representation is generated using a k-mer transformation: $$K: \Sigma^n \to (\Sigma^k)^{n-k+1}$$ such that for a sequence $s=s_1s_2 \dots s_n \in \Sigma^n$, $$\mathbf{K(s)} = [s_1\dots s_k, s_2\dots s_{k+1}, \dots s_{n-k+1}\dots s_{n}]$$The result is a vector of $n-k+1$ overlapping fixed length k-mers by a sliding window of stride one across the sequence. Overlapping k-mer sequence representation is widely used in computational biology for various sequence related tasks including classical sequence alignment algorithms and modern machine learning tasks \cite{ng2017dna2vec,Yanrong2021dnabert,ng2017dna2vec,orenstein2016kmers}.

\subsubsection{Tokenization}
The k-mer representation vector undergoes tokenization before it is fed to the encoder-decoder model. Tokenization is done using a word-piece tokenizer \cite{wu2016google} with vocabulary size of 261 tokens representing all 256 codewords in addition to the 5 special tokens: padding token, unused token, end-of-sentence token, classifier token, and a masking token. Note that while the tokenization deployment is done by word-piece tokenizer, the actual tokenization scheme represents a byte-level encoding, since every token represents a byte, much like \cite{xue2021byt5}.

\subsection{Architecture}
\begin{figure}[h]
   \centering
  \includegraphics[scale = 0.4]{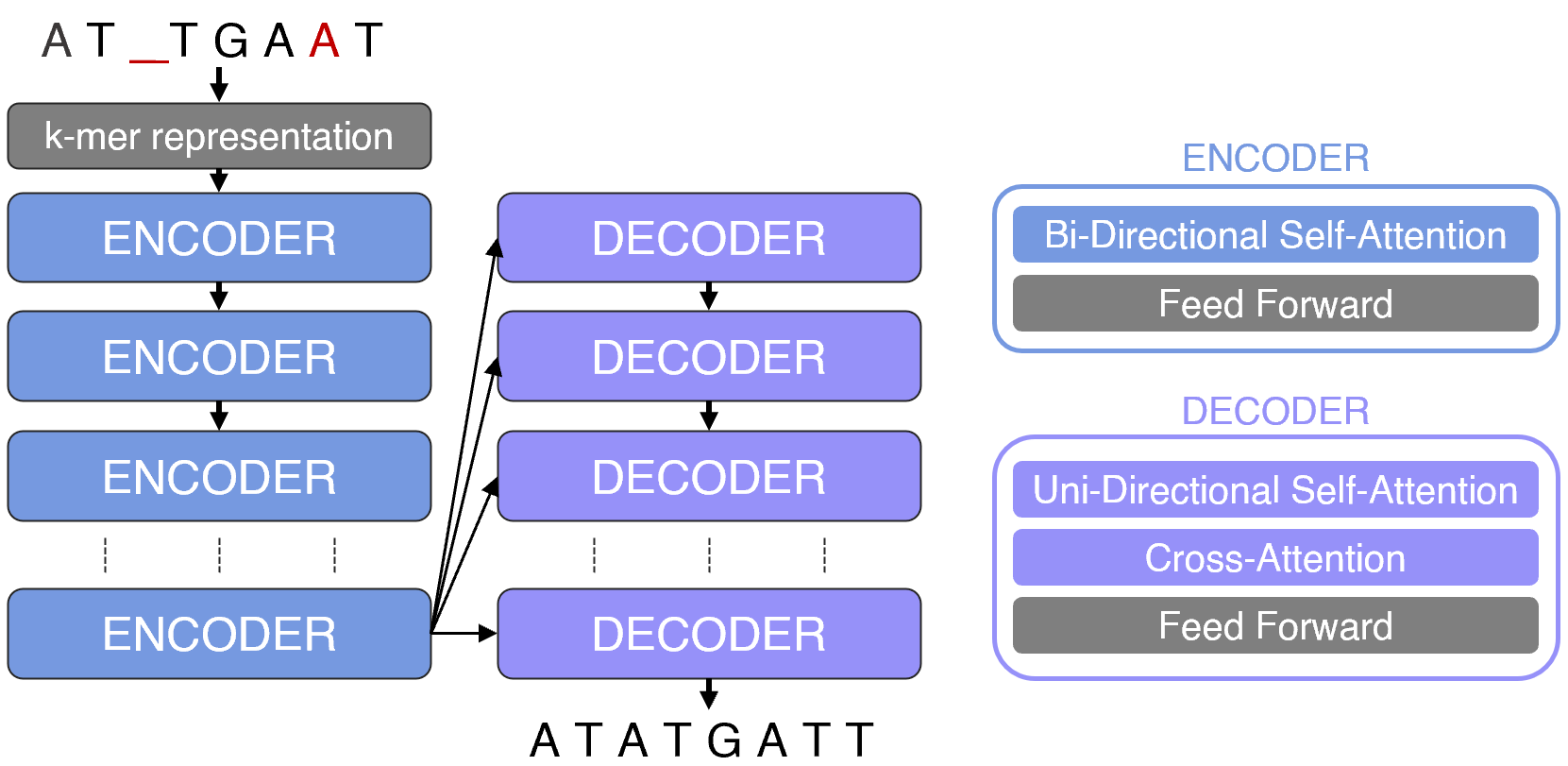}
   \caption{\textbf{Transformer Encoder-Decoder Architecture.}}
\label{fig:encoder}
\end{figure}

Figure \ref{fig:encoder} depicts the encoder-decoder Transformer composed of two paired BERT models \cite{devlin2018bert} used in this work. As suggested by \cite{rothe2020leveraging}, two main variations are made for the second BERT model, making it a decoder. First, in order to comply with auto-regressive generation, BERT’s bi-directional self-attention layers are changed to uni-directional self-attention layers. Second, since the decoder has to be conditioned on the contextualized encoded sequence, randomly cross-attention layers are added between the self-attention layer and the two feed-forward layers in each block. Apart from the aforementioned alterations, both models have the same configuration as the DNABERT 4-mer model from \citeauthor{Yanrong2021dnabert} with 12 layers, vocabulary size of 261, a hidden size of 768, filter size of 3072, and 12 attention heads. The weights are not shared between models, and are all randomly initialized. Note that this separation is necessary since the input for our model is a k-mer representation of the sequence, yet the output is the original sequence.
At every training stage, for a sequence of length $n$, the model is fed with it’s k-mer vector of size $n-k+1$. The model outputs a prediction sequence of constant size $l$, which is the length of the originally encoded sequence. During training, we use a cross-entropy loss between the decoder's output and the original input.

\begin{figure*}[h]
   \centering
  \includegraphics[scale = 0.55]{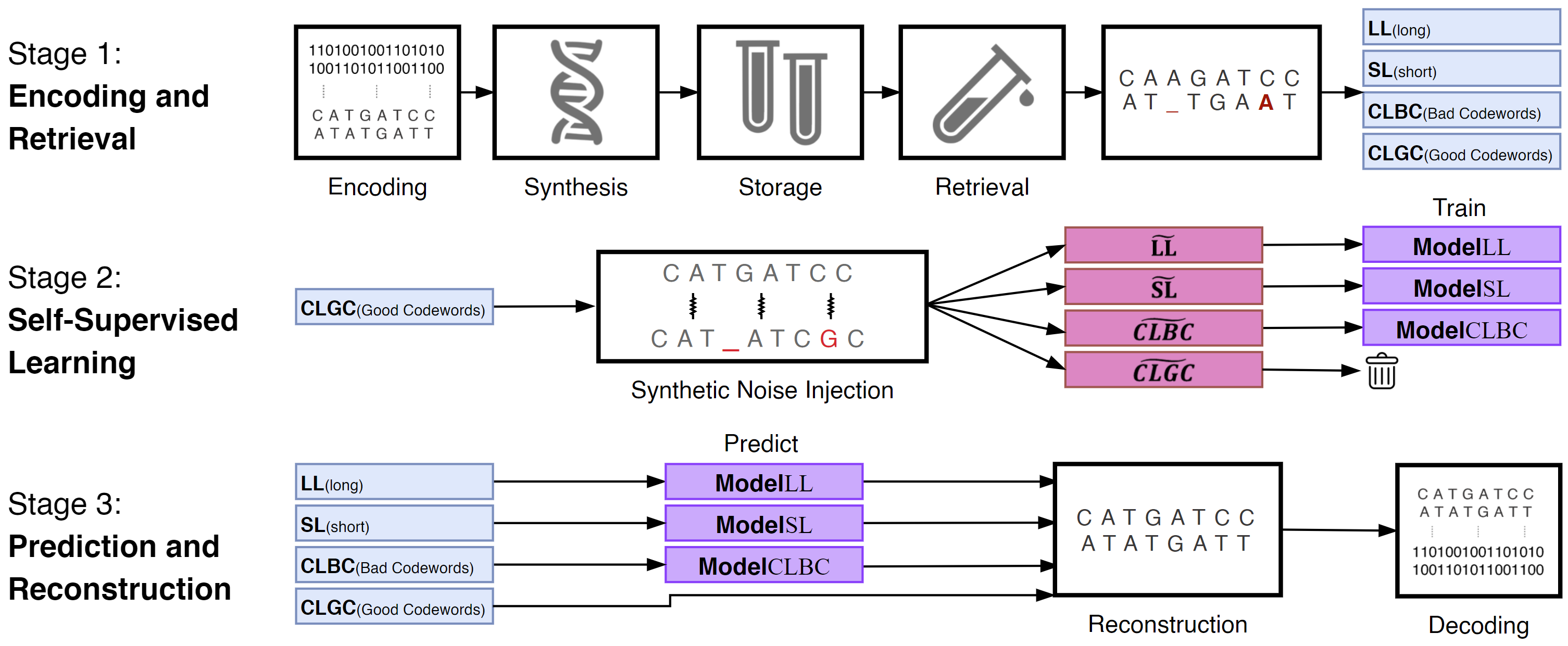}
  \caption{\textbf{Our proposed Single-Read Reconstruction Algorithm (SRR).} \textbf{Stage 1} includes byte-level encoding, DNA based storage and retrieval. \textbf{Stage 2} includes a rule based extraction of Correct Length Good Codewords ($CLGC$) reads as soft labeled data, synthetic noise injection producing $\widetilde{LL},\widetilde{SL},\widetilde{CLBC}$ that are used as training samples for training three reconstruction models. \textbf{Stage 3} includes reconstruction of the Long, Short and Erroneous reads ($LL$, $SL$, and $CLBC$ respectively) using constrained beam search predictions of the trained models. Finally, the original binary file is decoded.}
    \label{fig:overall_Method}
\end{figure*}

\section{Single-Read Reconstruction Using Transformers}
In this section we describe the end-to-end pipeline of our proposed Single-Read Reconstruction (SRR) scheme. The input to the pipeline is a given data file which undergoes encoding to DNA. After retrieval of the erroneous data sequences, we inject synthetic noise to a subset of ``quality" sequences, thus creating a training dataset for self-supervised learning which is tailor-made for the file. We use 3 trained models to predict and reconstruct the original data file. As presented in Figure \ref{fig:overall_Method}, our process is described in 3 main stages: \textbf{Stage 1:} Encoding to DNA and retrieval of noisy reads. \textbf{Stage 2:} Self-supervised training by synthetic noise injection. \textbf{Stage 3:} Data reconstruction using model prediction.

\subsection{Stage 1 - Encoding and Retrieval}
A given data file $m$ is encoded to DNA by mapping every 8-bits to a codeword of 4 DNA bases: $$E:\{0,1\}^8 \rightarrow \{A,C,G,T\}^4$$ yielding a set of 256 codewords. Namely, the encoding of the data file $m$ is given as: $$\hat{E}(m) = \{E(m_1m_2\dots m_8),E(m_9m_{10}\dots m_{16}),\dots\}$$This byte-level encoding fits commonly used data representations (bmp, UTF-8, etc.) and guarantees that the codewords also serve as semantic building blocks and thus hold meaningful information. This encoding provides us with a meaningful and expressive vocabulary on one hand (compared to a naive 2 bits per base encoding with only 4 codewords), but robust and generic that relies only on the byte-level encoding scheme, and is therefore compatible for various domains and file types. The complete DNA encoded data file is split to DNA sequences of a fixed length $l$. Together with the data, we store the set of codewords used by the encoder $E$ to encode the stored file. Note that the overhead for storing this set of valid codewords is limited by the number of available codewords $r\leq256$, which is negligible in large files.
The set of DNA sequences are then synthesized and stored. We retrieve noisy sequences with deletion, insertion and substitution errors.

\subsection{Stage 2 - Self-Supervised Learning}
\subsubsection{Noise Injection}
We divide the retrieved noisy reads into 3 sets: short sequences with length $<l$ ($SL$), long sequences with length $>l$ ($LL$) and sequences of a correct length $l$ ($CL$). Certainly, reads in $SL$ and $LL$ are the result of at least one deletion/insertion event, and possibly other errors as well. The $CL$ set is divided again into 2 subsets: sequences in which all codewords are valid ($CLGC$), meaning they contain only codewords that were used to encode the data file, and sequences containing invalid codewords ($CLBC$). Note that since we stored each data file together with the set of codewords used to encode it, we are able to distinguish the $CLGC$ from the $CLBC$ sets. Reads in $CLBC$ include some errors (most likely substitutions). The reads in $CLGC$ are treated as a set of sample sequences assumed to be correct and used to generate a training set for a self-supervised learning model. $t$ noisy copies of each sequence in the $CLGC$ set are produced with errors introduced with a random DIS channel. We again divide the noisy copies to 4 subsets: $\widetilde{SL}$, $\widetilde{LL}$, $\widetilde{CLBC}$, and $\widetilde{CLGC}$. We discard the $\widetilde{CLGC}$ set as it probably contains very few erroneous reads.

\subsubsection{Self-Supervised Training}
 The generated $\widetilde{SL}$, $\widetilde{LL}$, and $\widetilde{CLBC}$ sets are then pre-processed and used to train three different encoder-decoder models: Model$_{SL}$, Model$_{LL}$, and Model$_{CLBC}$ respectively, with the original $CLGC$ sequences as soft labels. The three models all share the same architecture described in the previous section. The term soft labels is due to the fact that the reads in $CLGC$ are not necessarily error-free. Intuitively, each of the three models is used to correct the error type most likely to occur in the corresponding training data.

\subsection{Stage 3 - Prediction and Reconstruction}
\subsubsection{Model Prediction}
Each of the 3 trained models, Model$_{SL}$, Model$_{LL}$, and Model$_{CLBC}$, is used to predict the corresponding sequence sets: $SL$, $LL$, and $CLBC$ respectively. Reads in $CLGC$ are assumed correct and are therefore predicted using the identity mapping.
\subsubsection{Constrained Beam Search}
Standard encoder-decoder models usually use beam search to improve model prediction quality. In a beam search strategy, every stage of the prediction saves the best $B$ (which is the beam size) candidates. We leverage the information regarding the expected output that is known from the encoding scheme to impose constraints to the basic beam search strategy. First, only output sequences of the correct length $l$ containing only valid codewords are allowed. In addition, out of the $B$ predicted output sequences, we select the one with the minimal edit distance to the erroneous input sequence.

\begin{table*}[h]
\begin{tabular}{ll|lll|lll|lll}
                             &                     & \multicolumn{3}{c|}{Darwin text} & \multicolumn{3}{c|}{Apollo code-script} & \multicolumn{3}{c}{Torres image}  \\
                             & Algorithm           & S        & E        & H          & S        & E        & H                 & S         & E        & H          \\ 
\hline
\multirow{2}{*}{Single copy} & Natural error       & $73.2\%$ & $0.34\%$ & $7.39\%$   & $72.6\%$ & $0.34\%$ & $7.60\%$          & $72.7\% $ & $0.34\%$ & $6.80\%$   \\
                             & SRR (\textit{Ours}) & $92.5\%$ & $0.10\%$ & $0.15\%$   & $91.9\%$ & $0.11\%$ & $0.13\%$          & $88.1\%$  & $0.18\%$ & $0.23\%$   \\ 
\hline
\multirow{2}{*}{2 copies}    & HRA                 & $85.7\%$ & $0.23\%$ & $0.64\%$   & $85.2\%$ & $0.25\%$ & $0.69\%$          & $87.9\%$  & $0.23\%$ & $0.71\%$   \\
                             & DBMA                & $84.6\%$ & $0.29\%$ & $0.83\%$   & $83.2\%$ & $0.32\%$ & $0.89\%$          & $87.4\%$  & $0.29\%$ & $0.72\%$   \\ 
\hline
\multirow{2}{*}{3 copies}    & HRA                 & $96.3\%$ & $0.07\%$ & $0.33\%$   & $96.3\%$ & $0.06\%$ & $0.29\%$          & $97.0\%$  & $0.05\%$ & $0.22\%$   \\
                             & DBMA                & $95.9\%$ & $0.09\%$ & $0.28\%$   & $95.8\%$ & $0.09\%$ & $0.27\%$          & $95.9\%$  & $0.09\%$ & $0.30\%$  
\end{tabular}

\caption{\textbf{Overall comparison of reconstruction algorithms on the three data files.} Success Rate (S), Mean Edit Error Rate (E), and Mean Hamming Error Rate (H) of the three reconstruction algorithms - our Single-Read Reconstruction Algorithm (SRR), Hybrid Reconstruction Algorithm (HRA) from \cite{Sabary:2020} and the Divider Bitwises Majority Alignment Algorithm (DBMA) which is an adjustment of the algorithm from \cite{BMA:18} done by \citeauthor{Sabary:2020}. SRR is tested on a single read while results of the other two algorithms are shown for 1,2, and 3 reads. For HRA and DBMA, Using a single read is equivalent to the natural error rate.}

\label{table:OverallPerformanceComparison}
\end{table*}

\section{Experimental Results}
To evaluate the performance of our reconstruction method, we performed three experiments with different types of data files. The first is a 948KB text file consisting of Darwin's ``On the Origin of Species" book \cite{darwin1909origin}. The second is a 3076KB text file containing the original code-script of the Apollo 11\footnote{Downloaded from https://github.com/chrislgarry/Apollo-11 on September 1, 2021}. This serves as a more complex case as the semantic structure of a code-script file is harder to identify than that of a text file from the English language. The third data file is a 4688KB color $1600\times1000$ image file of ``Torres del paine"\footnote{Image downloaded from https://www.travelandleisure.com on September 1, 2021}. The text and code-script files are UTF-8 encoded, and the image is encoded in the standard 8-bit color map bmp encoding. Namely, each pixel is represented by 3-bytes, one byte per color channel. We read the image file in $5\times5$ pixel squares. 

In all experiments, we assigned a codeword of 4 DNA bases to every byte of information, and encoded the data to DNA sequences of length $l=100$, each consisting of 25 encoded bytes.
The encoding resulted in 38,183, 125,977, 192,000 DNA sequences representing the Darwin, Apollo, and Torres files respectively. We simulated 1,2, and 3 noisy reads for each sequence using a simulator from \cite{dnasimulatorgadi} that introduces deletion, insertion and substitution errors according to the DIS channel. 

We reconstructed the original sequence from a single noisy read using our Single-Read Reconstruction Algorithm (SRR) and compared the result to the reconstruction results of running the Hybrid Reconstruction Algorithm (HRA) from \citeauthor{Sabary:2020} and their variation of the Divider Bitwise ajority Alignment Algorithm (DBMA) by \citeauthor{BMA:18} using one, two, and three reads. The implementations of HRA and DBMA are taken from \cite{dnasimulatorgadi,Sabary:2020}. Since both HRA and DBMA failed to run on the complete data files, the comparison is presented on a sample of 5,000 sequences from each file. For all algorithms, we evaluated success rates, edit error rates, Hamming error rates, and edit error distributions. For the noise injection step of the SRR algorithm we used error schemes representing standard DNA based storage conditions.  
Details of the experimental setting are available in the Technical Appendix.

\subsection{Overall Performance Comparisons}
To test the overall performance of the three algorithms we used the DNA channel simulator with an error scheme of synthesis by TWIST Bioscience and sequencing using Illumina MiSeq, representing standard conditions in DNA based storage studies \cite{organick2020probing,chandak2019improved,srinivasavaradhan2021trellis,anavy2019composite}. 

Table \ref{table:OverallPerformanceComparison} details the success rates, edit error rates and Hamming error rates for the three data files. Mean error rates are reported. For a detailed report including standard deviation, refer to the Technical Appendix. Evidently, our algorithm outperforms both HRA and DBMA using one and two reads in all three measures across the different data files. We reduced the mean natural edit error rates from $0.34\%$ to $0.1\%,0.11\%,0.18\%$ and successfully reconstructed $92.5\%,91.94\%,88.1\%$ of the sequences in the Darwin, Apollo, and Torres files respectively. The mean hamming error rates were reduced by $97\%$ from $7.26\%$ to $0.17\%$ on average on all files. We also achieve better results on the Hamming error rate compared to DBMA using three reads on all three files and on two out of the three files compared to HRA.

\begin{figure}[t]
    \centering
    \includegraphics[scale = 0.5]{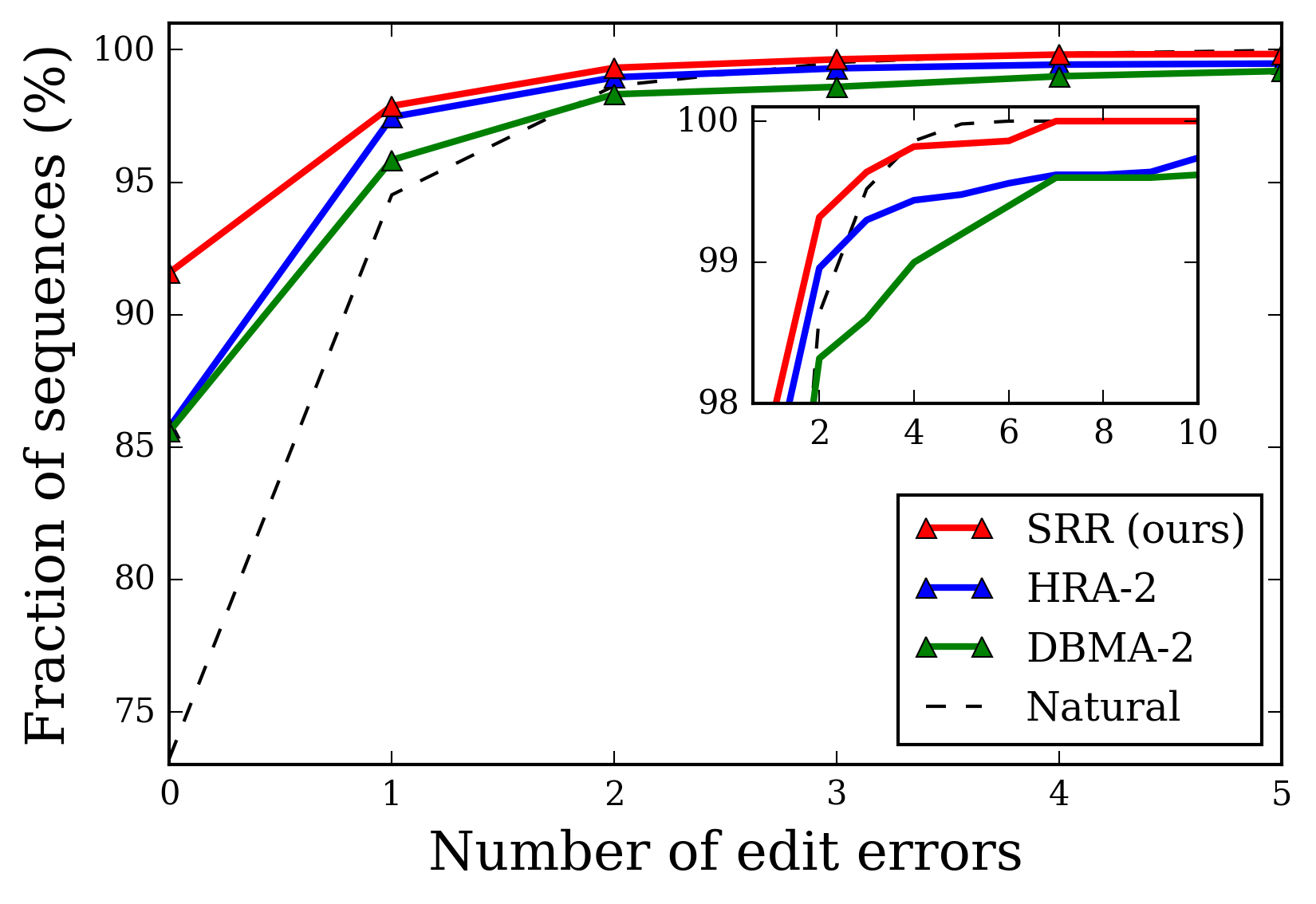}
    \caption{\textbf{Cumulative edit error histogram.} Edit errors frequencies tested in the Darwin text file using one and two reads. Presented as a cumulative histogram, the X-axis presents the number of edit errors while the Y-axis depicts the fraction of sequences with up to the number of edit error. The inset shows the distribution for the higher error values.}
    \label{fig:Cumm_Edit_Error}
\end{figure}

Figure \ref{fig:Cumm_Edit_Error} depicts the cumulative distribution of the edit errors of SRR, HRA using two reads and DBMA using two reads. Clearly SRR shows lower error rates. Moreover, while SRR saturates on reconstruction of 100\% of the sequences with up to seven edit errors, HRA and DBMA contain a long tail of a small number of sequences with up to 32 and 28 edit errors respectively. (See inset). This implies that when combined with error correction (EC) codes in an end-to-end DNA based data storage, SRR will require adding less redundancy in the EC scheme.

\subsection{Performance Comparison in High-Noise Regimes}
\begin{figure}[t]
    \centering
    \includegraphics[scale = 0.5]{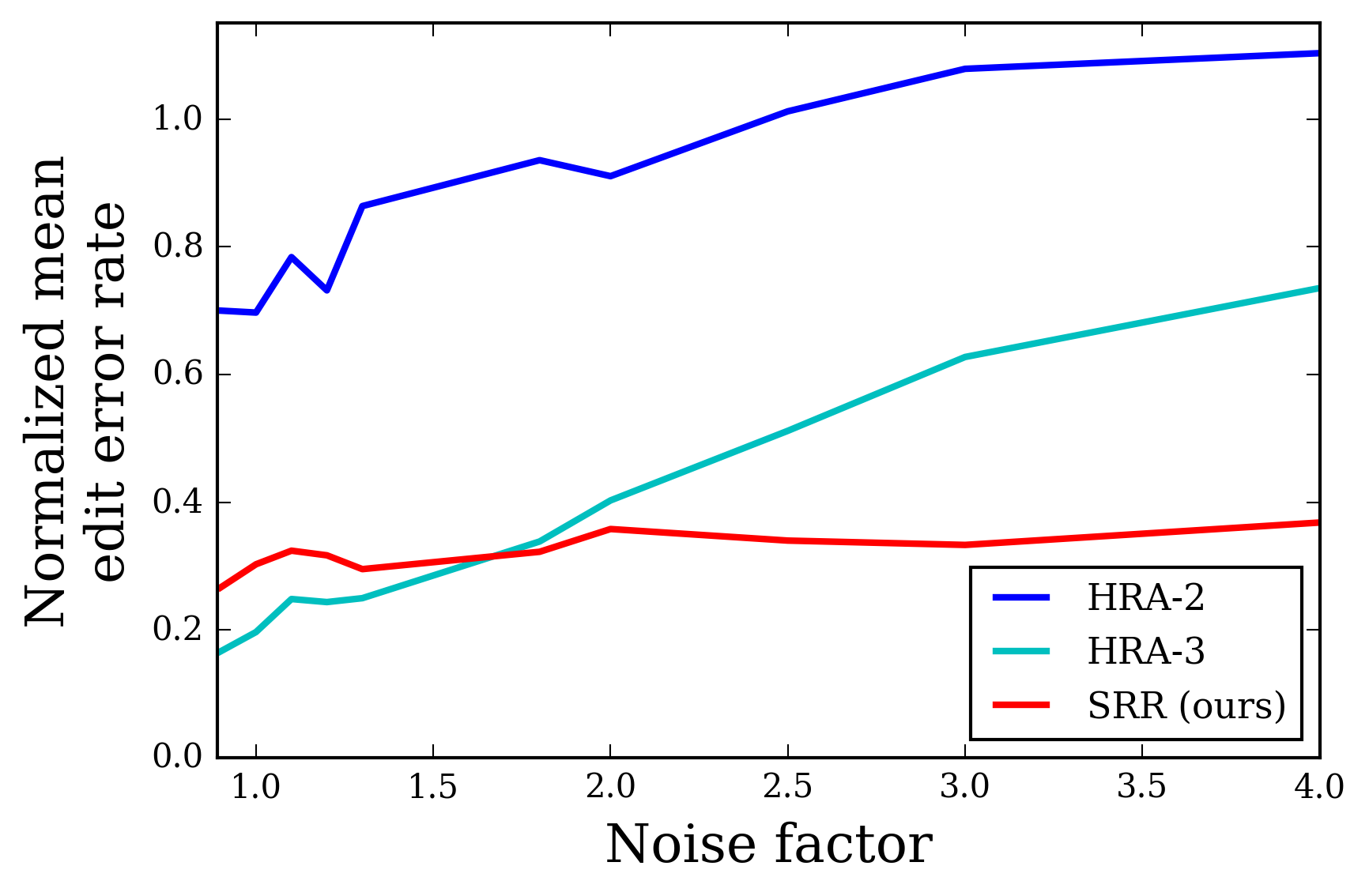}
    \caption{\textbf{The effect of increased noise levels.} Normalized mean edit error of HRA using two and three reads and of SRR as a function of the added noise factor.}
    \label{fig:Noise_Analysis}
\end{figure}

To assess the performance of our algorithm on data containing higher error rates we used the encoded Darwin text file and simulated input data with error rates that are $0.9-4$ times higher than the standard error rate in the previous experiment. Figure \ref{fig:Noise_Analysis} shows the mean edit error of SRR and HRA using two and three reads as a function of the added noise factor. The error rates are normalized by the natural error rate. While both HRA-2 and HRA-3 performance deteriorate as the noise level increases, SRR manages to reduce the natural error rate by a relatively constant fraction of about 65\%.

\begin{figure*}[h]
    \centering
    \includegraphics[scale = 0.4]{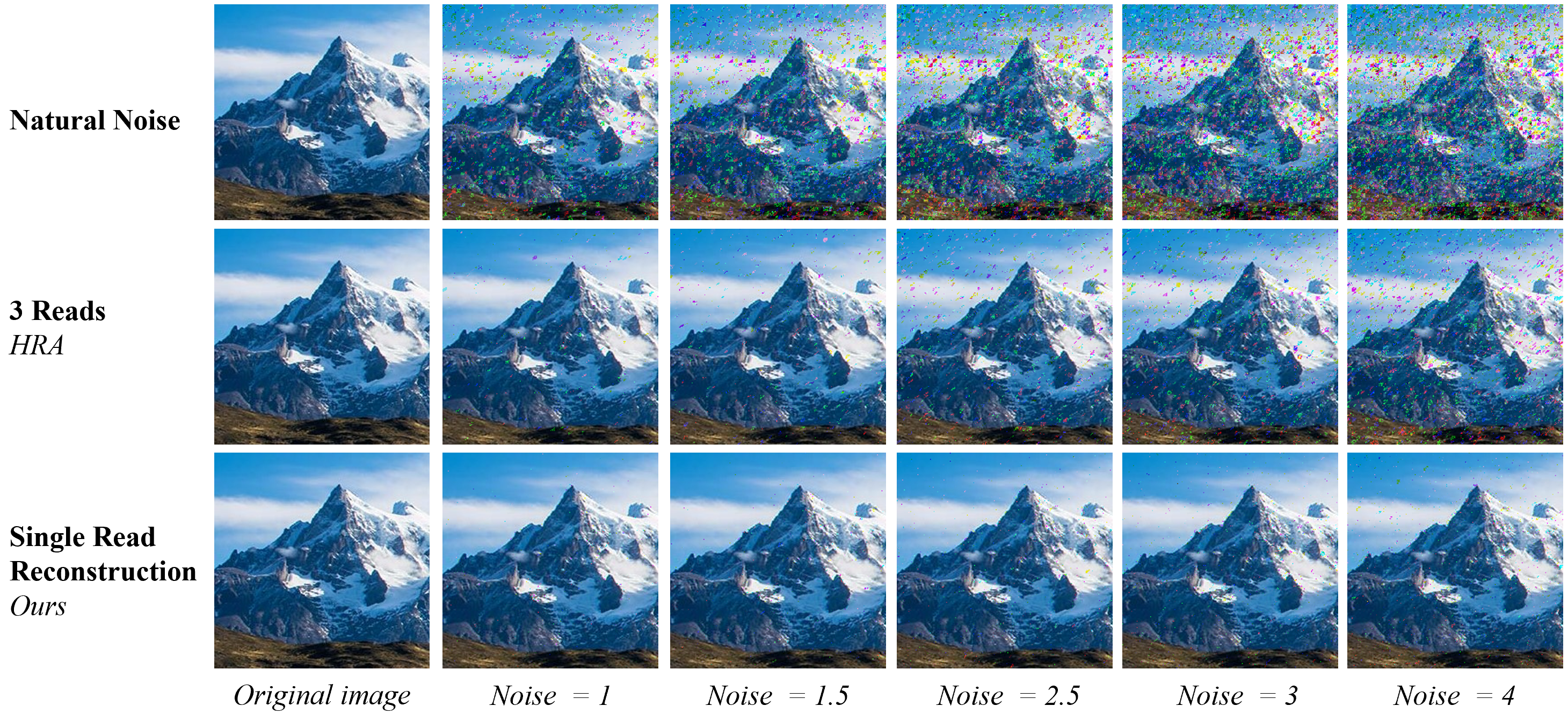}
    \caption{\textbf{The effect of increased noise levels.} Image reconstruction results of HRA using three reads and of SRR as a function of the added noise factor. Image downloaded from https://www.travelandleisure.com on September 1, 2021}
    \label{fig:Images}
    
\end{figure*}

To illustrate the effect of different reconstruction performance on data with different noise levels, we performed a similar experiment on the Torres del paine image file. The results of this experiment are visualized in Figure \ref{fig:Images}. This visualization clearly demonstrates the quality of the SRR algorithm. Even with a noise level four times higher than the standard error levels, the reconstruction using a single read is very successful while HRA with three reads suffer from high error rates.

\section{Discussion}
In this paper, we proposed a novel approach for utilizing deep learning Transformers to perform single-read reconstruction for DNA based storage. We modeled the error correction process as a sequence-to-sequence task and designed an encoder-decoder model to correct deletion, insertion, and substitution errors. Our approach is based on self-supervised training using only the decoded file with injected noise to produce training samples. We demonstrated our approach by encoding three file types and performing single-read reconstruction on simulated noisy data with different noise levels. Our single-read reconstruction algorithm achieves better performance than the current state-of-the-art reconstruction algorithms operating on two reads and in some measures also three reads. 

This is the first example of single-read reconstruction for DNA based storage. Traditional algorithms treat the reconstruction problem of each sequence independently and are thus limited when operating on single-read sequences. Our approach is the first, to our knowledge, to offer a reconstruction algorithm that is both \emph{holistic} (\textit{i.e.} reconstructs all sequences simultaneously) and \emph{context-aware}. Our suggested approach is thus able to successfully reconstruct single-read sequences. 

The common approach for using machine learning models for sequence reconstruction relies heavily on the availability of high-quality labeled data. This is used to try and learn the error pattern to allow the prediction of the original sequences. Dataset generation of synthetic DNA libraries requires extensive investments and labor and thus these datasets are scarce. Moreover, DNA synthesis and sequencing technologies, as well as the subsequent molecular biology procedures, produce diverse error patterns that are affected from various exogenous factors making the learning task implausible.  On the contrary, our model relies solely on self-supervised learning using noise injection on samples that are presumed to be error free. Our model does not require any external samples for training, making it applicable for use on diverse data files.

While we use a certain error distribution for noise injection, it does not need to represent the same error distribution as the decoded data file. This is demonstrated in the noise level experiment where our model outperformed the two and three reads reconstruction algorithms on data with noise levels much higher than those used for noise injection. We are able to achieve these results because our reconstruction model does not learn the error pattern in the data but rather it learns the \emph{underlying structure of the data file} in hand. Learning the underlying structure of the decoded file requires the identification of error-free reads that are assumed to be a representing sample of the original file. We use our knowledge of the error model (\textit{i.e.} errors are the result of some DIS channel) and a limited amount of redundant information added to the file in the encoding step (\textit{i.e.} the set of valid codewords in the file and the designed length $l$) to generate a set of reads that are error-free with high probability.

The possibility of obtaining a high quality set of probably-correct reads that adequately represent the input file is affected by three factors: file size, noise level, and the number of valid codewords in the file. Larger files allow for a larger and more diverse training set which is crucial for model training. Alternatively, high noise levels and larger sets of valid codewords result in a limited training set. An extreme case of the latter is demonstrated in the image encoding experiment as the set of valid codewords contains all 256 options. This  makes the distinction between correct and erroneous reads impossible and results in lower success rate and higher error rates compared to the text file experiments. Exploring ways to improve the construction of this read set is a major focus for future studies.

While most DNA based storage systems aim for an average read coverage of 10 reads or more, some sequences will inevitably be represented by only a single read. Salvaging those poorly represented sequences to extract more information will allow for the reduction of the overall coverage required for a successful decoding and the use of error correction schemes with less redundancy. Moreover, single-read reconstruction can be combined with standard multi-read reconstruction algorithms. For instance, after reconstructing each read using a single-read reconstruction model, reads can be clustered and used to reconstruct the sequence with higher confidence. 

In the past decade, the use of deep learning (DL) models has revolutionized many fields, including Computer Vision, Natural Language Processing, Cyber security, and others. Often, the introduction of DL models required rethinking central dogmas and traditional methodologies. DNA data storage is a fast growing field which gained a lot of attention in the past few years. However, despite its promise, the field seems to be limited in scale. This work represents a first step towards the incorporation of DL models in the DNA encoding-decoding pipelines. Similar to other domains, this may require a paradigm shift in the field. Replacing the traditional reconstruction algorithms and error correction schemes, that are focused on understanding the error patterns on the single sequence level, by a holistic, context-aware approach that tries to reconstruct the entire data file by modeling its underlying structure is a step in that direction.

\bibliography{aaai22}

\end{document}